\documentclass[% reprint,
 superscriptaddress,
 amsmath,amssymb,
 twocolumn,
 aps,
 pra,
 10pt,
longbibliography
]{revtex4-2}

\usepackage{float}
\usepackage{graphicx,xcolor}
\usepackage{enumitem}
\usepackage{braket}
\usepackage[colorlinks=true]{hyperref}
\usepackage{amsfonts,amssymb}
\usepackage{amsmath, mathtools}
\usepackage{caption,subcaption}

\usepackage[normalem]{ulem}

\usepackage{float}
\usepackage{graphicx,xcolor}
\usepackage{tikz}
\usetikzlibrary{arrows.meta}

\usetikzlibrary{positioning}
\usepackage[edges]{forest}
\usepackage{enumitem}
\usepackage{braket}
\usepackage{comment}

\usepackage{amsthm}

\newcommand{\Tr}{\rm Tr}
\def \unm {\mathbb{I}}

\def \dt {\text{d}}
\def \xr {\mathrm{x}}
\def \kr {\mathrm{k}}

\definecolor{amethyst}{rgb}{0.6, 0.4, 0.8}

\newcommand{\mbf}[1]{\mathbf{#1}}

\DeclarePairedDelimiter\abs{\lvert}{\rvert}

\begin{document}
\title{Harvesting magic from the vacuum}

\author{Ron Nystr\"om}
\email{ron.nystrom@helsinki.fi}
\address{Department of Physics, P.O.Box 64, FIN-00014 University of Helsinki, Finland}

\author{Nicola Pranzini}
\email{nicola.pranzini@helsinki.fi}
\address{Department of Physics, P.O.Box 64, FIN-00014 University of Helsinki, Finland}
\address{QTF Centre of Excellence, Department of Physics, University of Helsinki, P.O. Box 43, FI-00014 Helsinki, Finland}
\address{InstituteQ - the Finnish Quantum Institute, Finland}

\author{Esko Keski-Vakkuri}
\email{esko.keski-vakkuri@helsinki.fi}
\address{Department of Physics, P.O.Box 64, FIN-00014 University of Helsinki, Finland}
\address{InstituteQ - the Finnish Quantum Institute, Finland}
\address{Helsinki Institute of Physics, P.O.Box 64, FIN-00014 University of Helsinki, Finland}

\begin{abstract}
Magic is the quantum resource allowing a quantum computer to perform operations that cannot be simulated efficiently by classical computation. As such, generating magic in a quantum system is crucial for achieving quantum advantage. This letter shows that magic can be harvested by a three-level Unruh--DeWitt detector (a qutrit) interacting with a quantum field in an initial vacuum state. While the idea of extracting resources from Quantum Field Theories (QFT) was born from the harvesting of entanglement, our result extends the protocol to evolve a qutrit from a non-magical state to a magical one, making it possible to generate magic from QFT.
\end{abstract}

\maketitle

\textit{Introduction.--}  Magic is the quantum resource quantifying the ability of a quantum state to perform computational tasks that exceed the capabilities of classical systems~\cite{Gottesman:1998hu, BravyiK05}. Therefore, this resource is indispensable for achieving quantum advantage and is even more relevant than entanglement in quantum computing and the design of efficient quantum algorithms~\cite{VeitchETAl14, OlivieroEtAl2022}. Indeed, while entanglement describes non-classical correlations, i.e. quantifies the extent to which a state can be used for quantum information processing and communication protocols such as quantum teleportation, superdense coding, and quantum cryptography~\cite{Bennett98, HorodeckiEtAl09}, this resource alone does not quantify quantum advantage, since merely creating entanglement can be efficiently simulated classically~\cite{Gottesman:1998hu}.

In Relativistic Quantum Information (RQI), entanglement harvesting \footnote{``Harvesting'' is a standard name for the protocol, but note a difference in terminology: entanglement is not necessarily \textit{removed} from the vacuum, whereas e.g., harvesting corn from a field leaves few cobs left.} is the protocol that exploits a quantum field's vacuum state to generate entanglement between two causally disconnected non-relativistic quantum mechanical systems~\cite{Valentini91, Reznik03, MartinMartinezM14, Pozas-KerstjensEtAl15, SaltonElAt15, SuryaatmadjaEtAl22}. Despite being a relatively new concept, entanglement harvesting has already led to advancements in our understanding of the entanglement properties of Quantum Field Theory (QFT)~\cite{deS.L.TorresEtAl23, PercheEtAl24, AgulloEtAl23, AgulloEtAl24}. 
%\ekv{In this work, we ask what other interesting resources could be harvested.}
It is natural to ask if one can generate other qualitatively different quantum resources by interacting with a quantum field. This Letter takes a step forward by introducing a protocol for ``harvesting'' magic from the vacuum state of a QFT. Specifically, we investigate how coupling a non-relativistic three-level system, called a {\em detector}, to a generic quantum field can make the detector's state magical, demonstrating that the interaction with a field's vacuum can enhance the computational resources of a quantum system. As a proof of concept, we apply the framework to the case of a three-level system  (in quantum computation, a qutrit) coupled to a massless scalar field in flat spacetime, a paradigmatic example of what is known as an Unruh--DeWitt detector. By this setting, we find that the amount of harvested magic depends on the detector's internal structure, the details of the detector--field coupling, and the specific properties of the field's vacuum state, i.e. the values taken by one- and two-point functions.

In short, this letter introduces magic harvesting as a possible concept and outlines potential practical implementations and theoretical extensions.

\textit{Unruh--DeWitt detectors.---}
In RQI, %and the theory of Unruh--DeWitt detectors, 
an important model involves a non-relativistic quantum system - typically a two-level system (i.e. a qubit) - coupled to a quantum field, to explore the non-trivial notion of a vacuum state in non-inertial frames; in this context the system is called an Unruh--DeWitt detector after its pioneers~\cite{Unruh76, DeWitt80}. For this letter, it is more useful to consider a three-level system $\mathcal{D}$ with a free Hamiltonian
\begin{equation}
    \hat{H}_\mathcal{D}=\mathrm{diag}(0, \Omega_1, \Omega_1 + \Omega_2)~,
\end{equation}
and energy levels $\{\ket{0},\ket{1}, \ket{2}\}$ where $\Omega_1, \Omega_2$ are the first and second energy gap, respectively. This system moves in a $D$-dimensional spacetime along a trajectory parameterized via its proper time $\tau$ as $\xr(\tau)$ and interacts with a quantum field $\Phi$; following the standard convention, we call it an Unruh--DeWitt detector. Specifically, the detector--field coupling is given by the interaction picture Hamiltonian
\begin{equation}
    \hat{H}_{\text{int}}(\tau)=\lambda \chi(\tau)\hat{\mu}(\tau)\otimes\hat{O}_\Phi(\xr(\tau))
    \label{e.H}
\end{equation}
where $\lambda$ is an interaction-strength parameter, $\chi(\tau)$ is a switching function describing the range of times for which the detector and the field are allowed to interact, $\hat{\mu}$ is the (interaction picture) monopole moment operator
\begin{equation}
    \hat{\mu}(\tau) = \frac{1}{\sqrt{2}} \left( \ket{1}\bra{0} e^{i\Omega_1 \tau} + \ket{2}\bra{1} e^{i \Omega_2 \tau} \right) + \mathrm{h.c.}
\end{equation}
describing level transitions in the detector~\cite{LimaEtAl23}, and $\hat{O}_\Phi$ is some hermitian field operator evaluated along the trajectory of the detector. For the sake of simplicity, we require $\bra{\Phi_0}\hat{O}_\Phi\ket{\Phi_0}=0$, where $\ket{\Phi_0}$ is the vacuum state of the field. The above choice of Hamiltonian makes the interaction localized along the trajectory of the detector and provides the interaction picture joint time evolution operator
\begin{equation} \label{e.time_evol_U}
    \hat{U}_{\tau_i,\tau_f}=\mathcal{T}\left[\exp\left(-i\int_{\tau_i}^{\tau_f} d\tau \hat{H}_{\text{int}}(\tau)\right)\right]~,
\end{equation}
where $\mathcal{T}$ denotes the time-ordered product. Starting from the global state
\begin{equation}
    \ket{\Psi}=\ket{0}\otimes\ket{\Phi_0}~,
\end{equation}
the state of the detector after the interaction, as shown in App. \ref{a.rhoD}, is represented by the density operator
\begin{equation}
    \hat{\rho}_\mathcal{D}%={\rm Tr}_\phi[\hat{U}_{\tau_i,\tau_f}\ket{\Psi}\bra{\Psi}\hat{U}_{\tau_i,\tau_f}^\dagger]
    =\begin{bmatrix}
        ~p~ & 0 & \beta^* \\ ~0~ & q & 0 \\ \beta & 0 & 1-p-q
    \end{bmatrix}~,
    \label{e.rho}
\end{equation}
where the probability of finding one when measuring the system on the energy eigenbasis is
\begin{equation}
    q = \bra{1} \hat{\rho}_\mathcal{D} \ket{1} = \bra{\Phi_0} \hat{\mathcal{U}}^\dagger\hat{\mathcal{U}} \ket{\Phi_0}~,
\end{equation}
where $\hat{\mathcal{U}}=\bra{1}\hat{U}_{\tau_i,\tau_f}\ket{0}$ is some (possibly smeared) field operator completely determined by the interaction and trajectory above. In what follows, we will occasionally use quantum computation terminology, referring to the vectors $\{\ket{0},\ket{1},\ket{2}\}$ as the (qutrit's) computational basis. 

Assuming $\lambda$ to be small enough to allow the use of perturbation theory, we can find $q$ as
\begin{equation}
    q = \lambda^2 \mathcal{F}_q+\mathcal{O}(\lambda^4)
\end{equation}
where
\begin{equation}
      \mathcal{F}_q= \int_{\tau_i}^{\tau_f}d\tau \int_{\tau_i}^{\tau_f}d\tau' \chi(\tau)\chi(\tau')e^{-i\Omega_1(\tau-\tau')}\mathcal{O}(\tau,\tau')
\end{equation}
with 
\begin{equation}
    \mathcal{O}(\tau,\tau')= \bra{\Phi_0}\hat{O}_\Phi(\xr(\tau))\hat{O}_\Phi(\xr(\tau'))\ket{\Phi_0}~;
    \label{e.O}
\end{equation}
as we will see later, it is common to take $\hat{O}_\Phi$ as the field operator evaluated along the trajectory or some smeared version of it~\cite{Schlicht04}. Likewise, we obtain the off-diagonal element
\begin{equation}
    \beta = \bra{2}\hat{\rho}_\mathcal{D} \ket{0} =: \lambda^2 \mathcal{F}_\beta + \mathcal{O}(\lambda^4)
\end{equation}
with
\begin{equation}
\begin{split}
    \mathcal{F}_\beta = -\frac{1}{2} &\int_{\tau_i}^{\tau_f}d\tau \int_{\tau_i}^{\tau_f}d\tau' \chi(\tau)\chi(\tau') \\ \times \bigg[ &\Theta(\tau-\tau') e^{i(\Omega_1 \tau' + \Omega_2 \tau)} \mathcal{O}(\tau, \tau') \\ +~&\Theta(\tau'-\tau) e^{i(\Omega_1 \tau + \Omega_2 \tau')} \mathcal{O}(\tau', \tau) \bigg] 
\end{split}
\end{equation}
where the Heaviside step functions $\Theta$ appear by virtue of the time-ordering. 

We stress that, while necessary to obtain analytical and numerical results, the analysis in the next section does not rely on the use of perturbation theory: all our results can be obtained entirely in terms of $q$ an $\beta$, which is expressed exactly through \eqref{e.rho}.

\textit{Magic harvesting.---} 
%\textit{Magic.---} %\textcolor{red}{Magic is the resource describing the ability of a quantum system to perform tasks which cannot be efficiently simulated classically. Naively, one could think that entanglement cannot be simulated classically, but this is not the case~\cite{Gottesman98}. Indeed, Bell states are included in the so-called stabilizer set (${\rm STAB}$), which is obtained by the orbits of the Clifford group $\mbf{C}_n$, i.e. the normalizer of the Pauli group $\mbf{P}_n$~\cite{}. Therefore, quantum circuits made of gates from the Clifford group and starting from states on a computational basis can be simulated on a classical computer, regardless of whether or not they can produce entangled states. Consequently, non-classically simulable tasks can only be achieved by states not in ${\rm STAB}$. By definition, these can be obtained by applying non-Clifford gates, which thus represent a resource for quantum computing~\cite{OlivieroEtAl2022}.[...modify, add...]}
By Gottesman-Knill theorem, if a quantum algorithm does not contain non-Clifford operations, it can be simulated on a classical computer in polynomial time~\cite{Gottesman:1998hu}. Moreover, a universal set of gates must include a non-Clifford gate such as the T gate ($\pi/8$ gate) or the Toffoli gate \cite{Nielsen_Chuang_2010}.
Hence, non-Clifford gates are imperative for attaining quantum advantage in computation. Formally, the necessary non-Clifford gates belong to the set $\mathcal{C}_3 := \{ \hat{U} ~|~ \hat{U} \mathcal{C}_1 \hat{U}^\dagger \subseteq \mathcal{C}_2\}$, where $\mathcal{C}_2$ denotes the Clifford group, which is the normalizer of the Pauli group $\mathcal{C}_1$. Indeed, the set $\mathcal{C}_3$ includes both the Toffoli and T gates~\cite{GottesmanC99}. Alternatively, $\mathcal{C}_3$ gates can be replaced by Clifford gates and a supply of ancillary qubits in states containing a resource called non-Cliffordness or non-stabilizerness, which is often referred to as \emph{magic}~\cite{BravyiK05,VeitchETAl14}. Therefore, magic is the resource that describes the ability of a quantum system to perform tasks that cannot be efficiently simulated classically. Importantly, magic quantifies the advantage of quantum computation more fundamentally than entanglement, since highly entangled states may be classically simulable. For instance, the maximally entangled Bell states are contained in the stabilizer set, i.e. the orbit of $\mathcal{C}_2$, and thus contain zero magic~\cite{Gottesman:1998hu}. More generally, in quantum computing with qudits instead of qubits, the counterpart of requiring magic is asking for Wigner negativity, i.e. for states whose Wigner function in the discrete quantum phase-space takes negative values. Any computation involving only Wigner non-negative states can be efficiently simulated classically, while Wigner neqativity enables quantum advantage in universal computation \cite{Veitch_2012,Mari-Eisert,Veitch_2013}.

In this section, we propose a novel protocol to harvest magic from the vacuum state of a quantum field. More specifically, we show that the interaction Hamiltonian \eqref{e.H} can drive the state of the three-level system - a qutrit - to a state that cannot be efficiently simulated classically, i.e. has magic (i.e., Wigner negativity) \cite{Gross06,Gross07}. While magic is a basis-dependent concept, here there is a natural preferred basis: the energy eigenstates of the detector. Note that magic is not directly extracted from the field \footnote{For a discussion of whether the ground state of a field could have magic, see \cite{CFT-magic}.}; instead, the interaction with the field generates magic in the qutrit. To this end, we quantify magic by \emph{mana}, a quantity introduced in Ref.~\cite{VeitchETAl14}. By this measure, the magic of a mixed state $\hat{\rho}$ in some basis $\mathcal{B}$ is obtained as
\begin{equation}
    M(\hat{\rho}) := \ln \left[\sum_{a,a'=0}^{n-1} \abs{W_{(a,a')}(\hat{\rho})} \right]
\end{equation}
where
\begin{equation}
    W_{\mbf{a}}(\hat{\rho}) := n^{-1} \Tr(\hat{A}_{\mbf{a}}\hat{\rho})
\end{equation}
is the discrete Wigner function \cite{Wootters87, Gross06}, with $n$ denoting the dimensionality of the detector's Hilbert space and $\mbf{a}:= (a,a') \in \mathbb{Z}_n \times \mathbb{Z}_n$ labelling the points of a discrete phase-space. Note that the discrete Wigner function is defined only for odd $n$ \cite{FerrieE09}, which is why we focus our attention on qutrits rather than on the two-level systems naturally arising in RQI. As it is clear, computation of mana requires knowing the phase-space-dependent operators
\begin{equation}
    \hat{A}_{\mbf{0}} := n^{-1} \sum_{\mbf{a}} \hat{T}_{\mbf{a}}~, ~ \hat{A}_{\mbf{a}} := \hat{T}_{\mbf{a}} \hat{A}_{\mbf{0}} \hat{T}_{\mbf{a}}^\dagger
\end{equation}
given by the generalized Pauli matrices $T_\mbf{a}$, also known as the Weyl-Heisenberg matrices. These are generated by the shift and clock operators
\begin{equation}
    \hat{X} := \sum_{k=0}^{n-1} \ket{k+1~\mathrm{mod}~n}\bra{k}~, ~ \hat{Z}:= \sum_{k=0}^{n-1} \omega^k \ket{k} \bra{k}~,
\end{equation}
as
\begin{equation}
    \hat{T}_{aa'} = \omega^{-\frac{(n+1)}{2}aa'} \hat{Z}^a \hat{X}^{a'}~,
\end{equation}
where we employed a common shorthand notation for the phase $\omega := e^{2\pi i / n}$. Note that, as magic is basis-dependent, in this Letter we calculate mana in the interaction picture of the energy eigenbasis.

Using the density operator \eqref{e.rho}, and thus setting $n=3$, we obtain the magic for the detector as
\begin{widetext}
\begin{equation}
    M(\hat{\rho}) = \ln\left\{ 1-q + \frac{1}{3} \left[ \abs{q+2~\textrm{Re}(\beta)} + \abs{q-\textrm{Re}(\beta)-\sqrt{3}~\textrm{Im}(\beta)} + \abs{q-\textrm{Re}(\beta)+\sqrt{3}~\textrm{Im}(\beta)} \right] \right\}~.
\end{equation}
\end{widetext}
As it is clear, the state of the detector is mixed at the end of the protocol. While using mixed states to realize non-Clifford operations is possible, one often needs purer magical states for practical implementations. Magic state distillation was invented for this purpose. This method converts several copies of a mixed magical density operator into one magical state with higher purity~\cite{BravyiK05,Anwar_2012}. A possible topic for further investigation is to apply magic distillation to multiple copies of Unruh--DeWitt detectors prepared by the above protocol, to harvest a more practical resource for quantum applications.

\begin{figure}
    \centering
    \includegraphics[width=0.9\linewidth]{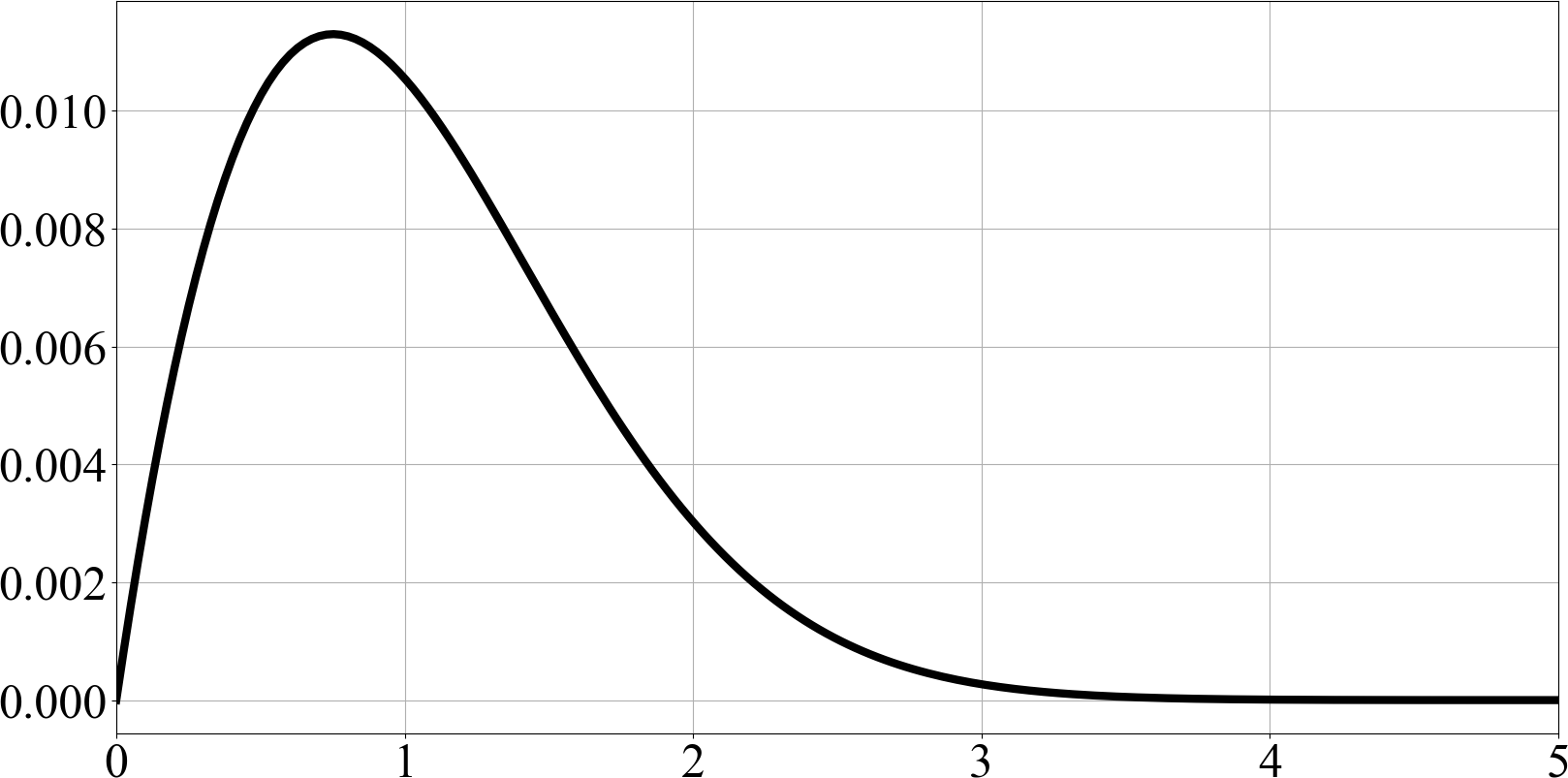}
    \caption{Mana $M/\lambda^2$ as function of the product of the energy gap and interaction length $\Omega \sigma_t$.}
    \label{f.Magic}
\end{figure}

\textit{Harvesting magic by an inertial point-like detector. --} As an example, in this section we compute the magic harvested by a point-like detector from the vacuum of a massless scalar quantum field. As proven above, we only need to compute $q$, i.e. the probability of finding the detector in $\ket{1}$, and $\beta$, i.e. the only off-diagonal element, after some interaction with the field takes place. To proceed further, we need to specify more details about the field state and the interaction. Specifically, we take the detector 
%with (interaction picture) monopole moment operator
%\begin{equation}
%    \hat{\mu}(\tau) = \ket{1}\bra{0} e^{i\Omega_1 \tau} + \ket{2}\bra{1} e^{i \Omega_2 \tau} + \mathrm{h.c.}
%\end{equation}
to move along an inertial trajectory $\xr (\tau)=(\tau, \mbf{x}_0)$, for which the massless scalar field operator reads
\begin{equation}
    \hat{\phi}(\xr) = \int \frac{\dt^D \kr }{(2\pi)^{D/2}} 
    \left( \hat{a}_{\mbf{k}}^\dagger e^{i \kr \cdot \xr} + \hat{a}_{\mbf{k}} e^{-i \kr \cdot \xr} \right)~,
\end{equation}
with roman letters $\xr$, $\kr$ denoting $D$-dimensional vectors in spacetime and $\hat{a}_{\mbf{k}}^{(\dagger)}$ being the annihilation (creation) operators in momentum space. For the sake of simplicity, we also take $D=4$, the detector to be point-like, the field to initially be in the Minkowski vacuum state $\ket{\phi_0}$, and the interaction to be described by a direct coupling between the monopole moment and the field operator. We also set the energy gaps to equal $\Omega:=\Omega_1=\Omega_2$ to avoid divergences. Thanks to these choices, Eq.~\eqref{e.H} becomes
\begin{equation}
    H_{\text{int}}(\tau)=\lambda \chi(\tau)\hat{\mu}(\tau)\otimes\hat{\phi}(\xr(\tau))~,
\end{equation}
and the expectation value \eqref{e.O} becomes the Wightman two-point function
\begin{equation}
    \mathcal{W}(\tau,\tau')= \bra{\phi_0}\hat{\phi}(\xr(\tau))\hat{\phi}(\xr(\tau'))\ket{\phi_0}~.
    \label{e.W}
\end{equation}
Finally, we must fix some switching function and the range of times for which the interaction is on. For the sake of simplicity, we pick $\chi(\tau)$ to be the unnormalised Gaussian
\begin{equation}
        \chi(\tau) = e^{-\tau^2 / \sigma_t^2}
\end{equation}
with variance $\sigma_t^2/2$. Considering contributions up to $\lambda^2$, we obtain the mana
\begin{equation}
    M(\hat{\rho}) = \frac{\lambda^2}{12 \sqrt{2\pi}} \Omega \sigma_t \left[ 1-\mathrm{erf}\left( \frac{\Omega \sigma_t}{\sqrt{2}} \right) \right] + \mathcal{O}(\lambda^4)~,
\end{equation}
which we plot for different values of $\sigma_t$ and $\Omega$ in Fig. \ref{f.Magic} (see App.~\ref{a.calc_q} and \ref{a.calc_b} for details). First, the magic harvested can be optimised by choice of energy gap $\Omega$ multiplied by the interaction length modulated by $\sigma_t$. 
%\sout{Second, the magic decreases as $\sigma_t$ becomes larger. We interpret this as a longer interaction time -- which is indeed effectively modulated by the magnitude of $\sigma_t$ -- and read it as the fact that the full system approaches a regime where its state is time-translation invariant, for which no excitation can occur: an inertial detector that is uniformly switched on for an infinite time gives a zero response regardless of the specific details of the interaction employed~\cite{Schlicht04}.} 
Second, fixing the energy gap, the amount of magic harvested increases with $\sigma_t$ only up to a certain point. After this point, increasing the interaction time will decrease the magic. We interpret this as the fact that the full system approaches a regime where its state is time-translation invariant, for which no excitation can occur: an inertial detector that is uniformly switched on for an infinite time gives a zero response regardless of the specific details of the interaction employed~\cite{Schlicht04}. Yet, the amount of harvested magic at finite times depends on the specific switching utilised. While using Gaussian smearing is customary in RQI, other functions may also be considered, and a bestiary of finite-time switchings has been presented in Ref.~\cite{SriramkumarP96}, where their extensions to infinite-time switching are also discussed. Hence, a more detailed analysis of the harvested magic vs shape and time of detector--field interaction is easily accessible; yet, we here only consider the Gaussian switching since the purpose of this example is only that of presenting a proof-of-concept for our magic harvesting protocol.

\textit{Conclusions.---}  With this letter, we presented a way to harvest magic from Quantum Field Theory (QFT) states. Our findings build on the principles of entanglement harvesting, showing these can be extended to magic and, therefore, can be used to extract computationally valuable resources from QFTs. In this sense, the implications of our results are far-reaching. On the one hand, our result may be implemented in real-life scenarios, providing costly quantum resources (i.e., non-Clifford operations) relatively inexpensively; based on this, Relativistic Quantum Information (RQI) may provide novel tools for performing non-classically simulable tasks. On the other hand, several extensions of our protocol to enhance magic harvesting are possible and involve exploring generic probes, diverse interactions, and other field theories. The following paragraphs explore these two potential outcomes of our results in more detail.

First, a compelling question is how to implement the harvesting protocol with real-life devices in the lab. One could speculate if in this way one could find an alternative method to provide magical resources to a quantum computer without requiring the design of non-Clifford gates. One might envision a quantum computer that exclusively performs Clifford operations but is supplemented with a collection of ancillary qubits or qudits coupled to a quantum field as proposed herein. Imagine such quantum computer performing Clifford operations freely while requesting magical ancillas only when needed. The ancillas are coupled with the field to extract magic and then included as a resource in the quantum registers used for the computation. An efficient application of this protocol might need more than one qubit to be simultaneously coupled to the field, hence generating many-qubit magical states; this scenario will be analyzed elsewhere. Another scenario could be to use the quantum field as a channel to transmit quantum information (magic) between the detectors \cite{Simidzija-etal, KasprzakT24}, i.e., in this case, for transmitting magic states. Discussion of possible concrete implementations of the transmission scenario, e.g. by coupling quantum dots to Luttinger liquids with potential experimental realizations, can be found in \cite{Aspling-etal}. Finally, the presented detector--field coupling may just be too cumbersome for experimental realisations of magic harvesting. As discussed above, we stress that magic is harvested from the interaction, not the environment; other types of interaction protocols and environments than the one discussed here may be more practical.

Second, several extensions of our proof-of-concept example are possible. A nontrivial setting mimicking entanglement harvesting would be to consider two spacelike separated detectors interacting with a field, and investigate whether non-local magic \cite{Cao2024nrx} could be generated into the joint state of the detectors.
Another question is whether one could construct a measure to directly quantify the magic in the time-evolution operator \eqref{e.time_evol_U}, in analogy of the Pauli instability developed to quantify magic in unitary circuits \cite{garcia2024}.
More straightforward extensions could consider different switching functions such as those mentioned in Ref.~\cite{SriramkumarP96}, focusing on the finite time interactions. This study would enable a selection of those switchings allowing a more efficient magic harvesting. Furthermore, the interaction Hamiltonian can be further generalized by considering different detector--field couplings or adding a finite extension of the detector in space; the latter can be achieved by using a standard procedure based on Fermi normal coordinates~\cite{Schlicht04, LoukoS06}. Considering different kinds of fields is also possible. This is in line with the the fact that, as mentioned above, we expect magic generation to be feasible in various settings involving a qudit interacting with an environment.

\textit{Acknowledgements.---} The authors thank Otto Veltheim for valuable comments and discussions, and Alan Robertson for a helpful lunch conversation. N.P. acknowledges financial support from the Magnus Ehrnrooth Foundation and the Academy of Finland via the Centre of Excellence program (Project No. 336810 and Project No. 336814). N.P. and E.K.-V. acknowledge the financial support of the Research Council of Finland through the Finnish Quantum Flagship project (358878, UH)

\bibliography{bib}

\newpage
\onecolumngrid
\appendix
\newpage
\section{Post-interaction density operator of detector}
\label{a.rhoD}
This appendix proves the form of the reduced density matrix $\hat{\rho}_\mathcal{D}$ presented in Eq. \eqref{e.rho}. Using the general Hamiltonian in Eq.~\eqref{e.H}, we show that some elements in the reduced density operator of the detector vanish. Let us denote the $m$-th term (of order $\lambda^m$) in the Dyson perturbative expansion of $\hat{U}$ as $\hat{U}^{(m)}$. Similarly, we denote
\begin{equation}
    \hat{\rho}^{(i,j)} := \Tr_\Phi [\hat{U}^{(i)} \hat{\rho}_0 \hat{U}^{\dagger (j)}]
\end{equation}
with $\hat{\rho}_0 = \ket{0}\bra{0} \otimes \ket{\Phi_0}\bra{\Phi_0}$ representing the initial state comprised of the detector in the ground state and field in the vacuum state, and $\hat{\rho}^{(i,j)}\sim \lambda^{i+j}$. The partial trace $\Tr_\Phi$ averages over all degrees of freedom in the field. Evidently, the state of the detector after the interaction becomes
\begin{equation}
    \hat{\rho}_\mathcal{D} = \sum_{i,j} \hat{\rho}^{(i,j)}
\end{equation}
up to some satisfactory order in $\lambda$.

We first notice that the $\hat{\mu}(\tau)$ operator (de-)excites the state of the detector s.t.
\begin{equation}
    \prod_{k=1}^m \hat{\mu}(\tau_k) \ket{0} = \begin{cases}
        c_0\ket{0} + c_2\ket{2} + c_4\ket{4} + \dots , \text{ if }m \text{ even} \\
        c_1\ket{1} + c_3\ket{3} + c_5\ket{5} + \dots , \text{ if }m \text{ odd}
    \end{cases}~,
\end{equation}
with $c_k \in \mathbb{C}$ denoting the relevant coefficients. This implies that to find the detector in the odd states, we must apply $U^{(k)}$ with odd $k$ on the initial state. The off-diagonal elements $\bra{m} \hat{\rho}^{(i,j)} \ket{l}$ must therefore vanish if $m+l$ is odd, i.e.
\begin{align}
\begin{split}
    \bra{m} \hat{\rho}^{(i,j)} \ket{l} \sim & \bra{m} \mathcal{T}\bigg[ \prod_{k=1}^i \hat{\mu}(\tau_k) \bigg] \ket{0}\bra{0} \mathcal{T}\bigg[ \prod_{k=i+1}^{i+j} \hat{\mu}(\tau_k) \bigg] \ket{l} \Tr_\Phi \left\{ \mathcal{T} \bigg[ \prod_{k=1}^{i} \hat{\mathcal{O}}_{\Phi}(\xr_k) \bigg] \ket{\phi_0}\bra{\phi_0} \mathcal{T} \bigg[ \prod_{k=i+1}^{i+j} \hat{\mathcal{O}}_{\Phi} (\xr_k) \bigg] \right\} \\
    \sim & \bra{\Phi_0} \mathcal{T} \bigg[ \prod_{k=1}^i \hat{\mathcal{O}}_{\Phi}(\xr_k) \bigg] \mathcal{T} \bigg[ \prod_{k=i+1}^{i+j} \hat{\mathcal{O}}_{\Phi}(\xr_k) \bigg] \ket{\Phi_0} \ \ \text{(with }i + j \text{ odd)}\\
        = & 0
\end{split}
\end{align}
This is zero $\forall i,j$ since we find $i+j$ to be odd, and hence there is an odd amount of field operators in the vacuum expectation value, which vanishes by Wick's theorem. Note that although the integrals were omitted, the resulting elements will vanish since the integrand evaluates to zero. As a consequence of this, setting the dimension amount of the detector to $n=3$, we find the general form of the reduced density matrix given by
\begin{equation}
    \hat{\rho} = \begin{bmatrix}
        p & 0 & \beta^* \\ 0 & q & 0 \\ \beta & 0 & 1-p-q
    \end{bmatrix}~,
\end{equation}
where we have required unit trace and hermiticity of the density operator.

\section{Explicit expression for $q$}
\label{a.calc_q}
This appendix shows the explicit procedure for obtaining $q$ up to $\lambda^2$. First, we expand the time evolution operator $\hat{U}_{\tau_i,\tau_f}$ in \eqref{e.time_evol_U} as a Dyson series, and denote with $\hat{U}^{(m)}$ the $m$-th term of the resulting infinite sum. In general, we denote the smearing function (describing both a switching in time and spatial smearing) as $\Lambda(\xr)$, and the initial state as $\hat{\rho}_0 = \ket{0}\bra{0} \otimes \ket{\phi_0}\bra{\phi_0}$. Considering a point-like detector (as in the main text), the general smearing above becomes a smooth switching function in time multiplied by a delta function in space. Taking $\tau_i \to -\infty, \tau_f \to \infty$ and applying perturbation theory yields
\begin{equation}
\begin{split}
    q &= \bra{1} \Tr_\phi [\hat{U}^{(1)} \hat{\rho}_0 \hat{U}^{\dagger (1)}] \ket{1} + \mathcal{O}(\lambda^4) \\
    &=\bra{1} \Tr_\phi \left[-i \int \dt \tau \hat{H}(\tau) \ket{0}\bra{0} \otimes \ket{\phi_0}\bra{\phi_0} i \int \dt \tau' \hat{H}(\tau') \right] \ket{1} + \mathcal{O}(\lambda^4) \\
    &= \lambda^2 \int \dt^D \xr \dt^D \xr' \Lambda(\xr) \Lambda(\xr') \bra{1} \hat{\mu}(\tau) \ket{0}\bra{0} \hat{\mu}(\tau') \ket{1} \Tr_\phi [ \hat{\phi}(\xr) \ket{\phi_0}\bra{\phi_0} \hat{\phi}(\xr ')] + \mathcal{O}(\lambda^4) \\
    &= \frac{\lambda^2}{2} \int \dt^D \xr \dt^D \xr' \Lambda(\xr) \Lambda(\xr') e^{i\Omega_1 (\tau-\tau')} \bra{\phi_0} \hat{\phi}(\xr') \hat{\phi}(\xr)\ket{\phi_0} + \mathcal{O}(\lambda^4) \\
    &= \frac{\lambda^2}{2} \int \dt^D \xr \dt^D \xr' \Lambda(\xr) \Lambda(\xr') e^{i\Omega_1 (\tau-\tau')} \mathcal{W}(\xr',\xr) + \mathcal{O}(\lambda^4) ~,
\end{split}
\end{equation}
where $\mathcal{W}(\xr',\xr)$ is the two-point Wightman function. For an inertial detector, this is evaluated as
\begin{equation}\label{e.W_calc}
\begin{split}
    \mathcal{W}(\xr_1, \xr_2) &= \int \frac{\dt^D \kr_1 }{(2\pi)^{D/2}} \int \frac{\dt^D \kr_2 }{(2\pi)^{D/2}} \bra{0} \bigg[ \hat{a}_{\mbf{k}_1}^\dagger e^{i \kr_1 \cdot \xr_1} + \hat{a}_{\mbf{k}_1} e^{-i \kr_1 \cdot \xr_1} \bigg] \bigg[ \hat{a}_{\mbf{k}_2}^\dagger e^{i \kr_2 \cdot \xr_2} + \hat{a}_{\mbf{k}_2} e^{-i \kr_2 \cdot \xr_2} \bigg] \ket{0} \\
    & \to \int \frac{\dt^d \mbf{k}_1 \dt^d \mbf{k}_2}{(2\pi)^{d} \sqrt{2\abs{\mbf{k}_{1}} 2\abs{\mbf{k}_{2}}}} \bra{0} \hat{a}_{\mbf{k}_{1}} \hat{a}_{\mbf{k}_{2}}^\dagger e^{i\kr_2 \cdot \xr_2 - i\kr_1 \cdot \xr_1} \ket{0} \\
    &= \int \frac{\dt^d \mbf{k}_1 \dt^d \mbf{k}_2}{(2\pi)^d 2 \sqrt{\abs{\mbf{k}_1} \abs{\mbf{k}_2}}} e^{i(-\abs{\mbf{k}_1}\tau_1 + \mbf{k}_1\cdot\mbf{x}_1 + \abs{\mbf{k}_2}\tau_2 - \mbf{k}_2 \cdot \mbf{x}_2)} \delta^{(d)}(\mbf{k}_1 - \mbf{k}_2) \\
    &= \int \frac{\dt^d \mbf{k}}{(2\pi)^d 2 \abs{\mbf{k}}} e^{-i\abs{\mbf{k}} (\tau_1 - \tau_2) + i\mbf{k}\cdot( \mbf{x}_1 -\mbf{x}_2 )}~,
\end{split}
\end{equation}
where we inserted the conditions of positive energy and zero mass to find
\begin{equation}
    \int \frac{\dt^D \kr}{(2\pi)^{D/2}} \sqrt{2\pi} \delta(\kr^2) \theta(\kr^0) = \int \frac{\dt^d \mbf{k}}{\sqrt{(2\pi)^d 2 \omega_{\mbf{k}}}}~,
\end{equation}
and the energy of the massless scalar particle is $\omega_{\mbf{k}} = \abs{\mbf{k}}$. In the last row of Eq. (\ref{e.W_calc}) we renamed $\mbf{k}:=\mbf{k}_1$ for simplicity.

Next, we focus on the case of a point-like detector moving in spacetime along an inertial trajectory $\xr_0(\tau)=(\tau,\mbf{x}_0)$, and smear the interaction in time according to the unnormalised Gaussian
\begin{equation}
    \Lambda(\tau,\mbf{x}) := \delta^{(d)}(\mbf{x}-\mbf{x}_0) e^{-\tau^2 / \sigma_t^2}~.
\end{equation}
We then evaluate $q$ by
\begin{equation}
\begin{split}
    q &= \frac{\lambda^2}{2} \int \dt^D \xr \dt^D \xr' \Lambda(\xr) \Lambda(\xr') e^{i\Omega_1 (\tau-\tau')} \mathcal{W}(\xr',\xr) \\
    &= \frac{\lambda^2}{2} \int \dt \tau \dt \tau' e^{-(\tau^2 + \tau'^2) / \sigma_t^2} e^{i\Omega_1 (\tau-\tau')} \int \dt^d \mbf{x} \dt^d \mbf{x}' \delta^{(d)}(\mbf{x}-\mbf{x}_0) \delta^{(d)}(\mbf{x}'-\mbf{x}_0) \mathcal{W}(\xr', \xr) \\
    &= \frac{\lambda^2}{2} \int \dt \tau \dt \tau' e^{-(\tau^2 + \tau'^2)/\sigma_t^2} e^{i\Omega_1 (\tau-\tau')} \int \frac{\dt^d \mbf{k}}{(2\pi)^d 2 \abs{\mbf{k}}}  e^{i\abs{\mbf{k}} (\tau - \tau')} \\
    &= \frac{\lambda^2}{2} \int \dt \tau \dt \tau' e^{-(\tau^2 + \tau'^2)/\sigma_t^2} e^{i\Omega_1 (\tau-\tau')} \frac{2\pi^{d/2}}{\Gamma(\frac{d}{2})} \frac{1}{2 (2\pi)^d } \int_0^\infty \dt \abs{\mbf{k}} \abs{\mbf{k}}^{d-2}  e^{i\abs{\mbf{k}} (\tau - \tau')}~,
\end{split}
\end{equation}
and regularise this by shifting $\tau \to \tau + i \varepsilon$ for some real, arbitrarily small $\varepsilon > 0$. In this way, we find
\begin{align}
\begin{split}
    \int_0^\infty \dt \abs{\mbf{k}} \abs{\mbf{k}}^{d-2}  e^{i\abs{\mbf{k}} (\tau - \tau')} &\to \int_0^\infty \dt \abs{\mbf{k}} \abs{\mbf{k}}^{d-2}  e^{i\abs{\mbf{k}} (\tau - \tau')} e^{-\abs{\mbf{k}} \varepsilon}   \\
    &= \frac{\Gamma(d-1)}{[\varepsilon - i(\tau - \tau')]^{d-1}} \ \ \text{(if } d > 1 \text{ )} \\
    &\to \frac{\Gamma(d-1)}{[-i(\tau-\tau')]^{d-1}}~,
\end{split}    
\end{align}
taking $\varepsilon \to 0$ in the last line, which leaves us with
\begin{equation}
    q = \frac{\lambda^2}{2} \frac{\Gamma(d-1)}{(4\pi)^{d/2} \Gamma(\frac{d}{2})} \int \dt \tau \dt \tau' \frac{1}{[-i(\tau-\tau')]^{d-1}} e^{-(\tau^2 + \tau'^2)/\sigma_t^2} e^{i\Omega_1(\tau-\tau')}
\end{equation}
To perform this intergal, we make the substitutions $s := \tau - \tau'$ and $u := \tau$ in the region $\tau > \tau'$ (lower triangle in the $(\tau,\tau')$-plane), and $s := \tau' - \tau$ and $u := \tau'$ in the region $\tau < \tau'$ (upper triangle in the $(\tau,\tau')$-plane). To integrate over all times $\tau, \tau'$, we sum the results we get from integrating over these two regions. As the integral over $s$ runs from $0$ to $+\infty$, we can swap the $s$ and $u$ integrals to obtain
\begin{align}
\begin{split}
    q &= \frac{\lambda^2}{2} \frac{\Gamma(d-1)}{(4\pi)^{d/2}\Gamma(\frac{d}{2})} \int_0^\infty \dt s \int_{-\infty}^\infty \dt u e^{-(s^2 + 2u^2 - 2su)/\sigma_t^2} \left[ \frac{1}{[-is]^{d-1}} e^{i\Omega_1 s} + \frac{1}{[is]^{d-1}} e^{-i\Omega_1 s} \right] \\
    &= - \frac{\lambda^2}{2} \frac{\Gamma(d-1)}{(4\pi)^{d/2}\Gamma(\frac{d}{2})} \sqrt{2\pi} \sigma_t \left[ \int_0^\infty \dt s \frac{1}{s^{d-1}} \sin(\frac{d\pi}{2}+\Omega_1 s)e^{-s^2/(2\sigma_t^2)} \right] \\ 
    &= \frac{\lambda^2}{8\pi} \left\{ e^{- \frac{1}{2} \Omega_1^2 \sigma_t^2} - \Omega_1 \sigma_t \sqrt{\frac{\pi}{2}} \left[ 1 - {\rm erf} \left( \frac{\Omega_1 \sigma_t}{\sqrt{2}} \right) \right] \right\} + \mathcal{O}(\varepsilon)~,
\end{split}
\end{align}
where we employed dimensional regularisation, took $d= 3-\varepsilon$ (with $D=d+1$), and expanded at zeroth order in $\varepsilon$. This is the same result obtained in Sec.~3.1 of \cite{SriramkumarP96} with an extra factor $1/2$ coming from normalisation of $\hat{\mu}(\tau)$.

\section{Explicit expression for $\beta$} \label{a.calc_b}
This appendix presents the derivation for obtaining the off-diagonal element $\beta$ up to $\lambda^2$. Similar to the previous appendix, we use the Dyson series to find
%\begin{alignat}
%    \beta =& \bra{2} \Tr_\phi [U^{(2)} \rho_0 \unm] \ket{0} \\ \notag
%    =& -\frac{\lambda^2}{2} \int \dt^D \xr \dt^D \xr' && \Lambda(\xr)\Lambda(\xr') \bra{2}\mathcal{T}\left\{ \mu(\tau) \mu(\tau') \right\} \ket{0} \bra{\phi_0} \mathcal{T}\{ \phi(\xr) \phi(\xr') \} \ket{\phi_0} \\ \notag
%    =& -\frac{\lambda^2}{2} \int \dt \tau \dt \tau' &&\bigg[ \Theta(\tau-\tau') e^{-(\tau^2 + \tau'^2)/\sigma_t^2} e^{i\Omega_2 \tau + i \Omega_1 \tau'} \mathcal{W}(\tau,\tau') \\ \notag
%    & &&+ \Theta(\tau'-\tau) e^{-(\tau^2 + \tau'^2)/\sigma_t^2} e^{i\Omega_1 \tau + i \Omega_2 \tau'} \mathcal{W}(\tau',\tau) \bigg] \\
%    =& -\frac{\lambda^2}{2} \frac{\Gamma(d-1)}{(4\pi)^{d/2} \Gamma(\frac{d}{2})} \int \dt \tau \dt \tau' && \bigg[ \Theta(\tau-\tau') e^{-(\tau^2 + \tau'^2)/\sigma_t^2} e^{i\Omega_2 \tau + i \Omega_1 \tau'} [i(\tau-\tau')]^{1-d} \\ \notag
%    & &&+ \Theta(\tau'-\tau) e^{-(\tau^2 + \tau'^2)/\sigma_t^2} e^{i\Omega_1 \tau + i \Omega_2 \tau'} [-i(\tau-\tau')]^{1-d} \bigg] \notag
%\end{alignat}
\begin{align}
    \beta =& \bra{2} \Tr_\phi [U^{(2)} \rho_0 \unm] \ket{0} \notag \\ \notag
    =& -\frac{\lambda^2}{2} \int \dt^D \xr \dt^D \xr' \Lambda(\xr)\Lambda(\xr') \bra{2}\mathcal{T}\left\{ \mu(\tau) \mu(\tau') \right\} \ket{0} \bra{\phi_0} \mathcal{T}\{ \phi(\xr) \phi(\xr') \} \ket{\phi_0} \\ \notag
    =& -\frac{\lambda^2}{4} \int \dt \tau \dt \tau' \bigg[ \Theta(\tau-\tau') e^{-(\tau^2 + \tau'^2)/\sigma_t^2} e^{i\Omega_2 \tau + i \Omega_1 \tau'} \Tilde{\mathcal{W}}(\tau,\tau') + \Theta(\tau'-\tau) e^{-(\tau^2 + \tau'^2)/\sigma_t^2} e^{i\Omega_1 \tau + i \Omega_2 \tau'} \Tilde{\mathcal{W}}(\tau',\tau) \bigg] \\
    =& -\frac{\lambda^2}{4} \frac{\Gamma(d-1)}{(4\pi)^{d/2} \Gamma(\frac{d}{2})} \int \dt \tau \dt \tau' \bigg[ \Theta(\tau-\tau') e^{-(\tau^2 + \tau'^2)/\sigma_t^2} e^{i\Omega_2 \tau + i \Omega_1 \tau'} [i(\tau-\tau')]^{1-d} + (~\tau\leftrightarrow \tau'~) \bigg] ~ .
\end{align}
Here we use the shorthand notation
\begin{equation}
    \Tilde{\mathcal{W}}(\tau, \tau') := \frac{\Gamma(d-1)}{(4\pi)^{d/2} \Gamma(\frac{d}{2})} [i(\tau - \tau')]^{1-d}
\end{equation}
to denote the Wightman function for a point-like detector at point $\mbf{x}_0$.

Next, we apply the substitution $x:=\tau-\tau', y:=\tau+\tau'$ s.t. $\tau = (x+y)/2, \tau'=(y-x)/2$ and Jacobian is $1/2$. We find
\begin{alignat}{2}
    \beta =& -\frac{\lambda^2}{4} \frac{\Gamma(d-1)}{(4\pi)^{d/2} \Gamma(\frac{d}{2})} \frac{1}{2} \int_{-\infty}^\infty \dt y  && \int_{-\infty}^\infty \dt x\bigg[ \Theta(x) e^{-(x^2 + y^2)/(2\sigma_t^2)} e^{i[\Omega_2(x+y) + \Omega_1 (y-x)]/2} [ix]^{1-d} \notag \\
    & &&+ \Theta(-x) e^{-(x^2 + y^2)/(2\sigma_t^2)} e^{i[\Omega_1 (x+y) + \Omega_2 (y-x)]/2} [-ix]^{1-d} \bigg] \notag \\
    =&-\frac{\lambda^2}{8} \frac{\Gamma(d-1)}{(4\pi)^{d/2} \Gamma(\frac{d}{2})} \int_{-\infty}^\infty \dt y && \bigg[ \int_{0}^\infty \dt x e^{-(x^2 + y^2)/(2\sigma_t^2)} e^{i[\Omega_2(x+y) + \Omega_1 (y-x)]/2} [ix]^{1-d} \notag \\
    & &&+ \int_{-\infty}^0 \dt x e^{-(x^2 + y^2)/(2\sigma_t^2)} e^{i[\Omega_1 (x+y) + \Omega_2 (y-x)]/2} [-ix]^{1-d} \bigg]  \\
    =&-\frac{\lambda^2}{8} \frac{\Gamma(d-1)}{(4\pi)^{d/2} \Gamma(\frac{d}{2})} \int_{-\infty}^\infty \dt y && \int_{0}^\infty \dt x \bigg[ e^{-(x^2 + y^2)/(2\sigma_t^2)} e^{i[\Omega_2(x+y) + \Omega_1 (y-x)]/2} [ix]^{1-d} \notag\\
    & && + e^{-(x^2 + y^2)/(2\sigma_t^2)} e^{i[\Omega_1 (y-x) + \Omega_2 (x+y)]/2} [ix]^{1-d} \bigg] \notag\\
    =&-\frac{2\lambda^2}{8} \frac{\Gamma(d-1)}{(4\pi)^{d/2} \Gamma(\frac{d}{2})} \int_{0}^\infty \dt x && \int_{-\infty}^\infty \dt y e^{-(x^2 + y^2)/(2\sigma_t^2)} e^{i[\Omega_2(x+y) + \Omega_1 (y-x)]/2} [ix]^{1-d}~, \notag
\end{alignat}
where we set $x \to -x$ in integral over $(-\infty, 0]$. To avoid the divergence caused by having different gaps in the detector, we set $\Omega := \Omega_1=\Omega_2$ and obtain
\begin{equation}
\begin{split}
    \beta =&- \frac{\lambda^2}{4} \frac{\Gamma(d-1)}{(4\pi)^{d/2} \Gamma(\frac{d}{2})} \sqrt{2\pi} \sigma_t e^{-\Omega^2\sigma_t^2/2} \int_{0}^\infty \dt x e^{-x^2/(2\sigma_t^2)} [ix]^{1-d} \\
    =& - \frac{\lambda^2}{4} \frac{\Gamma(d-1)}{(4\pi)^{d/2} \Gamma(\frac{d}{2})} \sqrt{2\pi} \sigma_t^2 (i\sigma_t)^{1-d} \Gamma(1-\frac{d}{2}) e^{-\Omega^2\sigma_t^2/2} \\
    =& - \frac{\lambda^2}{16\pi} e^{-\frac{1}{2}\Omega^2 \sigma^2}
\end{split}
\end{equation}
where we set $d=3$ in the last line.
\end{document}